Response to Solicitation NNH12ZDA008L: Science Objectives and Requirements for the Next
NASA UV/Visible Astrophysics Mission Concepts

Title: The History of Star Formation in Galaxies

Authors: Thomas M. Brown (Space Telescope Science Institute; tbrown@stsci.edu)
         Marc Postman (Space Telescope Science Institute; postman@stsci.edu)
         Daniela Calzetti (University of Massachusetts; calzetti@astro.umass.edu)

Point of contact: Thomas M. Brown
Position: Astronomer and JWST Mission Scientist
Organization: Space Telescope Science Institute
Email: tbrown@stsci.edu
Phone: 410-338-4902
Willingness to participate and present at NASA workshop: Yes

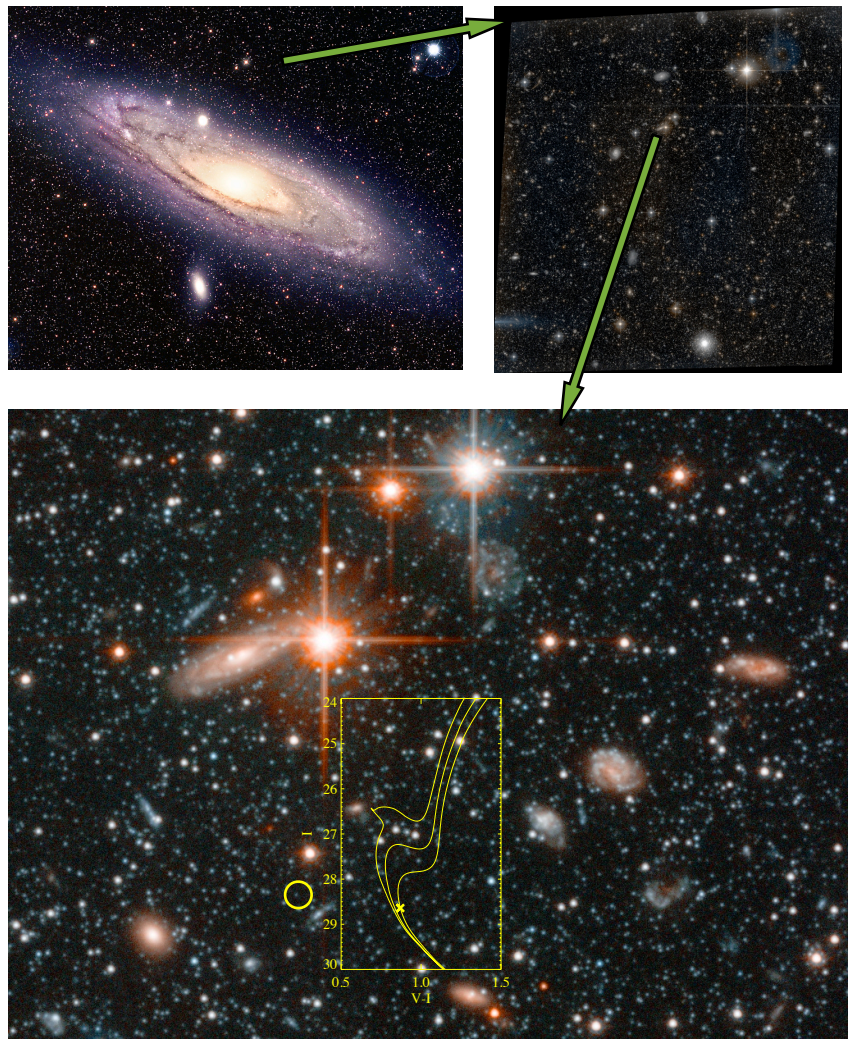

## Abstract

If we are to develop a comprehensive and predictive theory of galaxy formation and evolution, it is essential that we obtain an accurate assessment of how and when galaxies assemble their stellar populations, and how this assembly varies with environment. There is strong observational support for the hierarchical assembly of galaxies, but our insight into this assembly comes from sifting through the resolved field populations of the surviving galaxies we see today, in order to reconstruct their star formation histories, chemical evolution, and kinematics. To obtain the detailed distribution of stellar ages and metallicities over the entire life of a galaxy, one needs multi-band photometry reaching solar-luminosity main sequence stars. The *Hubble Space Telescope* can obtain such data in the low-density regions of Local Group galaxies. To perform these essential studies for a fair sample of the Local Universe, we will require observational capabilities that allow us to extend the study of resolved stellar populations to much larger galaxy samples that span the full range of galaxy morphologies, while also enabling the study of the more crowded regions of relatively nearby galaxies. With such capabilities in hand, we will reveal the detailed history of star formation and chemical evolution in the universe.

## Introduction

The study of galaxy evolution is pursued on two distinct fronts: the high-redshift universe and the local galactic neighborhood. In the high-*z* universe, we are directly observing the evolution of galaxies with time in an enormous sample of galaxies, but the properties of interest (morphology, kinematics, age, metallicity) are not directly accessible; instead, crude and degenerate diagnostics can be used in a composite sense on the scale of a resolution element. In the local volume, we can use methods that accurately measure these properties in resolved stellar populations, but we can only do so in a small sample of galaxies as they presently exist. High-redshift and galactic neighborhood studies are clearly complementary, but together they do not yet adequately explore the required parameter space. Progress in high-*z* work will require pushing back to the time of first light and the birth of galaxies, while progress in the local volume requires we accurately measure star formation histories beyond the Local Group.

The most robust method for measuring the star formation history of a stellar population comes from analysis of a color-magnitude diagram (CMD) that includes both the bright giant stars and the faint dwarf stars. Today, the *Hubble Space Telscope (HST)* can obtain the detailed star formation history in populations within a Mpc, but unfortunately our immediate neighborhood is too rural for such work (see van den Bergh 2000). The Local Group is a cosmological backwater, with only two giant spiral galaxies (the Milky Way and Andromeda) and a few dozen dwarf galaxies (mostly in orbit around the two giants). Some of the major morphological classes are not represented at all (e.g., giant elliptical and lenticular galaxies), and even the classes that are represented are not present in statistically meaningful numbers. For example, the intermediate-mass spiral M33 is the most common type of spiral in the universe (Marinoni et al. 1999), but M33 is the only representative of such a system in the Local Group. Beyond the Milky Way system, we have obtained a handful of *HST* pencil beams through some Local Group galaxies, but we simply have not characterized the assembly history in a representative sample of galaxies nor done so over a representative range of substructures of the known galaxy classes. The result is that our understanding of the star formation history in galaxies is highly skewed toward the few accessible examples (in the case of spirals) or based upon indirect and degenerate diagnostics (in the case of ellipticals, which are currently too distant for direct methods). Even so, the small steps we have taken so far have changed the way we view the assembly of galaxies, and obtaining a fair sample of stellar populations in galaxies is assured to yield substantial breakthroughs.

We shall use the recent exploration of Andromeda (M31) as an example. Out to a distance of ~25 kpc in the M31 halo, the metallicity exceeds that in our own halo by an order of magnitude (Mould & Kristian 1986; Durrell et al. 1994, 2001). These studies obtained the metallicity distribution from the colors of red giant branch (RGB) and asymptotic giant branch (AGB) stars. In principle, the luminosity distributions of the RGB and AGB stars also provide insight into the age of a population, in broad age bins of young (< 3 Gyr), intermediate age (3 - 8 Gyr), and old (8 - 13 Gyr) stars. In practice, obtaining these luminosity distributions is difficult to accomplish in sparse field populations, due to the combination of photometric scatter, broad metallicity range, uncertainties in apparent distance modulus, poor statistics for the brightest giants, and contamination from foreground stars and background galaxies. As a consequence, the M31 halo was assumed to be ancient (> 10 Gyr old). That picture changed when *HST* was able to image faint main sequence stars in M31 (albeit outside of the crowded interior). The M31 halo was found to host significant numbers of intermediate-age stars, presumably from a significant merger event (Brown et al. 2003; Fig. 1). A followup program probed the giant tidal stream and outer disk of M31 with main sequence photometry. The metallicity and age distributions in the stream were found to be very similar to those in the halo, suggesting that the halo was polluted with the debris from this disrupted satellite (Brown et al. 2006). This hypothesis was further borne out by N-body simulations (Fardal et al. 2007) and kinematic surveys (Gilbert et al. 2007). The population in the outer disk of M31 appears to be similar to that in the local Galactic thick disk, and does not include as many young stars as some disk formation models predict (Brown et al. 2006). Around the same time, an extended metal-poor halo was found in M31 (Guhathakurta et al. 2005; Irwin et al. 2005; Kalirai et al. 2006), spanning 20 degrees on the sky. There was speculation that this outer halo was the "true" halo, perhaps being both metal-poor and ancient, but deep photometric programs found intermediate-age stars in the extended halo as well (Brown et al. 2008). While semi-analytical models of galaxy formation have been used to simulate merger histories for the giant galaxies (e.g., Bullock & Johnson 2005), they have not made firm predictions on the distributions of age and metallicity in the disrupted satellites. These star formation history investigations and others like them are starting to provide the data required to constrain the populations in these hierarchical assembly histories (e.g., Font et al. 2008). Ironically, we know more about the age distribution in the M31 halo than we do in the Milky Way halo, due to reddening and distance uncertainties in the latter. There is some indication the Galaxy has had an unusually quiescent merger history, and that M31 is more representative of giant spiral galaxies (e.g., Hammer et al. 2007), but there is no way to know the variety of star formation histories in galaxies without a significant sample to explore.

The reconstruction of star formation histories in a wide range of galaxy types will reveal not only their star formation histories, but also the types of galaxies that were hierarchically assembled to construct the galaxies we see today.  Despite the enormous successes of the cold dark matter (CDM) paradigm, CDM predicts that giant galaxies such as the Milky Way and M31 should be surrounded by many more dwarf galaxies than actually observed (e.g., Moore et al. 1999).   As one way of solving the missing satellite problem, theorists have suggested that most dark matter sub-halos form few or no stars, due to the reionization of the universe (e.g., Tumlinson 2010; Bovill & Ricotti 2009).  The discovery of tidally disrupted satellites around the Milky Way (e.g., Sgr dwarf; Ibata et al. 1994) and M31 (e.g., the giant stellar stream; Ibata et al. 2001) rekindled searches for faint or disrupted satellites in the Local Group that would possibly account for these missing satellites.  Large surveys such as the Sloan Digital Sky Survey have discovered many new members of the Milky Way (e.g., Willman et al. 2005; Zucker et al. 2006) and M31 (e.g., Zucker et al. 2007; Majewski et al. 2007) systems, including the new class of ultra-faint dwarf galaxies.  Recent *HST* observations of the ultra-faint dwarfs have found they are the only known galaxies comprised solely of ancient stars (> 13 Gyr old), implying their star formation histories were truncated by reionization, and lending credence to the idea that most dark matter sub-halos remain dark (Brown et al. 2012).   However, no surveys will ever find the satellites that were already dispersed during hierarchical assembly.  Insight into these objects must come from analysis of the field populations in the giant galaxies of today.  Indeed, the population in the inner halo of M31 has been tied to a specific merger event that has not yet completed, as described above.

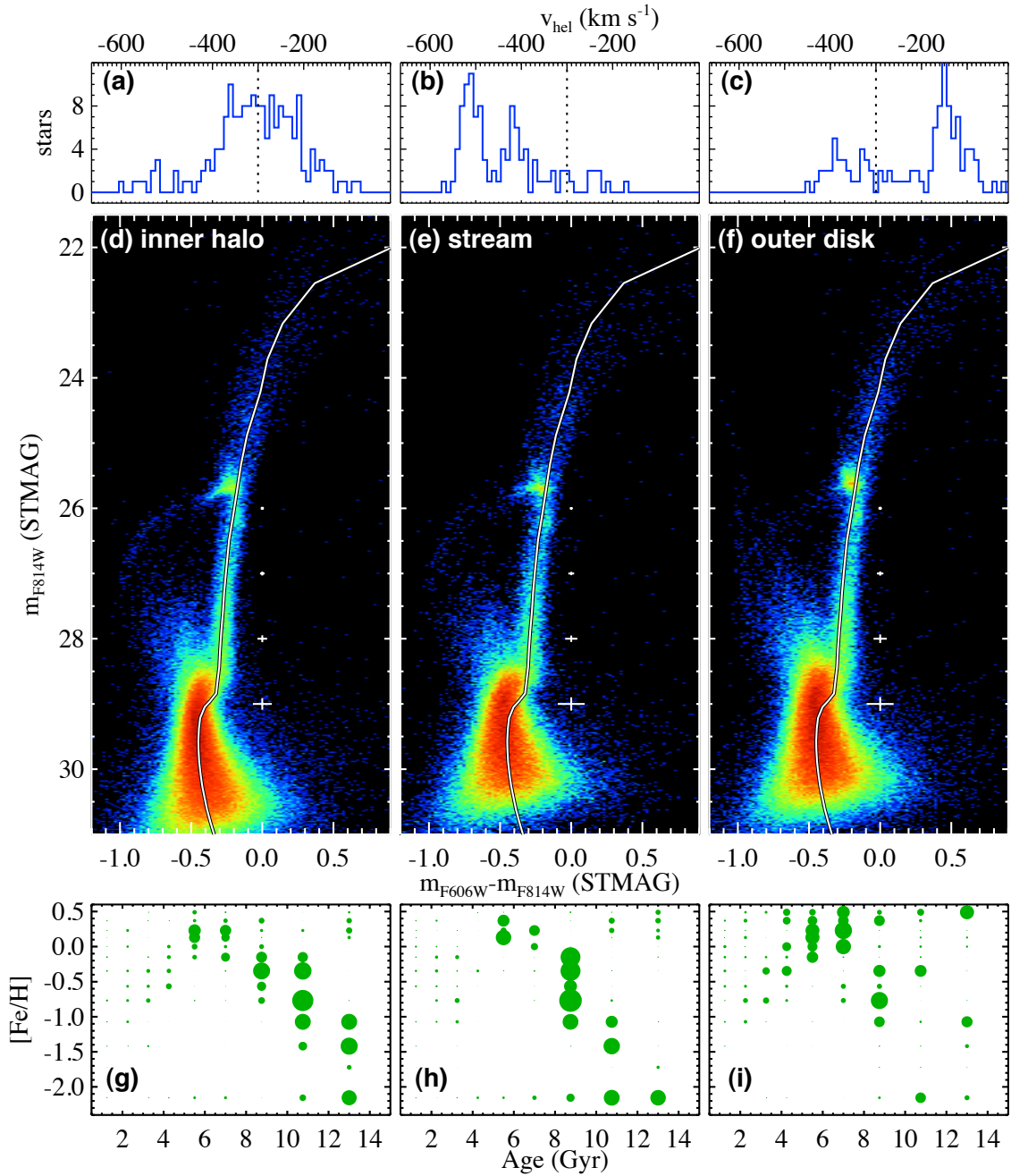

**Fig. 1 -** A reconstruction of the star formation histories in various structures of M31 (Brown et al. 2006). *(a-c):* Radial velocities obtained with Keck for fields in the inner halo (11 kpc on the minor axis), tidal debris stream (20 kpc off-axis), and outer disk (25 kpc on the major axis), with M31 systemic velocity indicated (*dotted line*). *(d-f):* CMDs in these structures, constructed from *HST* images reaching *V*~30 mag, with a ridge line of the globular cluster 47 Tuc (white *curve*) shown for comparison. *(g-i):* Star formation history in each field, with the area of the circles proportional to the weight in the fit. The inner halo and tidal stream each show a similar history of extended star formation, due to the debris from the stream polluting the inner halo. The outer disk has a population similar to that in the thick disk of the solar neighborhood.

## Methodology

The best tool for the reconstruction of star formation histories in nearby galaxies is photometry reaching dwarf stars on the main sequence.  The color and luminosity of the main sequence turn-off and subgiant branch are very sensitive to both metallicity and age, while the color of the RGB is much more sensitive to metallicity than age.  Spectroscopy of the bright giants can provide additional metallicity constraints (total metallicity, alpha enhancement, etc.) and kinematic information.  With a CMD that includes both the bright giant stars and faint dwarf stars, one can disentangle the effects of age and metallicity to obtain the detailed distribution of these parameters in a stellar population (Fig. 2).  A CMD that achieves a signal-to-noise ratio (SNR) of 5 at a point ~0.5 mag below the oldest main sequence turnoff in a population allows the reconstruction of the star formation history with age bins of ~1 Gyr over the entire lifetime of a galaxy, and metallicity bins of ~0.2 dex over the full range of abundances.  A solar analog (absolute $M_V$~5 mag) is a familiar and approximate reference point for the depth that must be reached.  The fitting of such CMDs was originally restricted to Galactic star clusters (e.g., Sandage 1953), but over time the fitting of CMDs has expanded to cover composite populations, first in Galactic satellites but eventually throughout the Local Group (e.g., Tosi et al. 1991; Gallart et al. 1999; Holtzman et al. 1999; Harris & Zaritsky 2001; Dolphin 2002; Brown et al. 2006, 2012; Cole et al. 2007).

For background-limited observations with a diffraction-limited telescope, the distance at which one can obtain photometry of faint stars is linearly proportional to the aperture diameter, assuming all other parameters are held fixed (bandpass, exposure time, SNR, instrument performance, etc.).  This is true in both sparse and crowding-limited regions.  In M31, *HST* can obtain

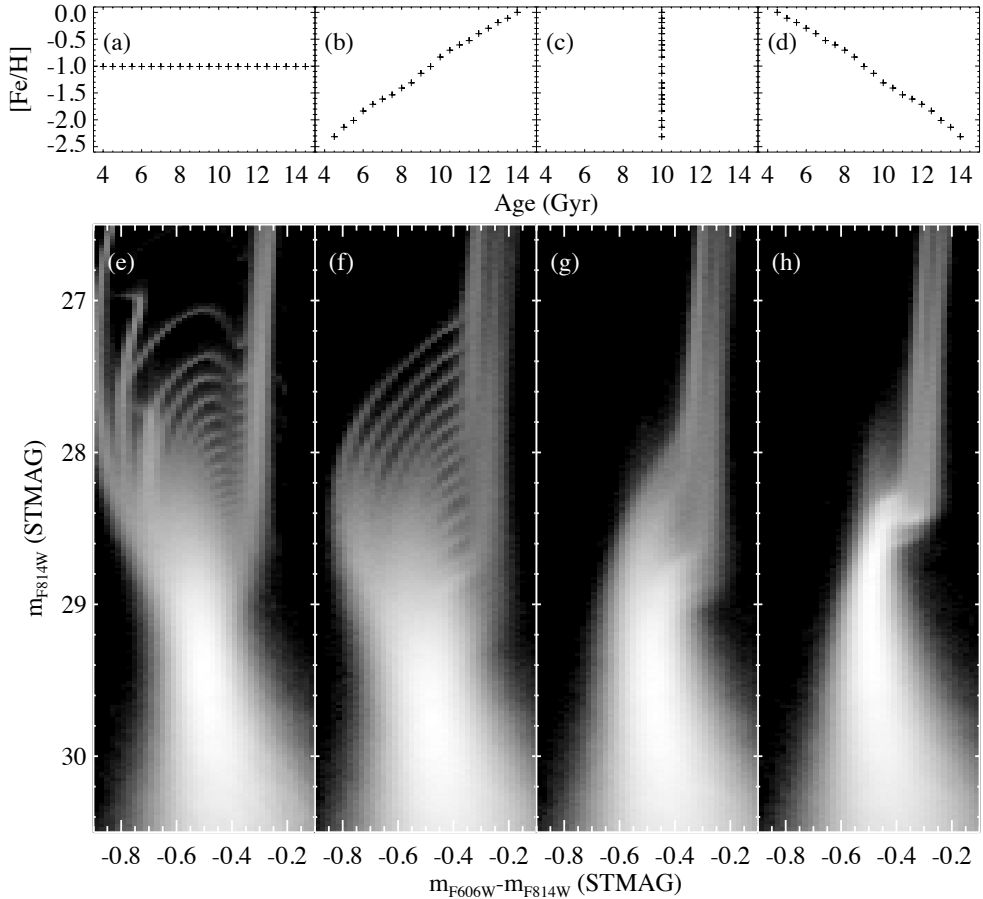

**Fig. 2 -** *Top panels:* Four hypothetical populations of stars. In each population, the stars are equally distributed among 20 isochrones distinct in age and metallicity. *Bottom panels:* Model CMDs for these hypothetical populations, with the observational errors of observations obtained in the M31 halo using *HST* (Brown et al. 2006). Because the main sequence turnoff, subgiant branch, and red giant branch all respond differently to changes in age and metallicity, a CMD that includes both the faint dwarfs and bright giants in a population breaks the age-metallicity degeneracy that would be present in observations of stars in a single evolutionary stage.

photometry of faint main sequence stars in regions where the surface brightness is roughly 26 V mag arcsec$^{-2}$ or fainter. This brightness falls at ~10 kpc on the minor axis and ~25 kpc on the major axis; the interior is currently unavailable to such probes. Although the field cannot be too crowded, it cannot be too sparse, either, because an accurate star formation history in a complex population requires a CMD of ~10,000 stars (for age bins of ~1 Gyr, metallicity bins of ~0.2 dex, and sensitivity to sub-populations at the ~20% level). To do analogous work in a galaxy 10 times further away than M31, we need a telescope with an aperture that is 10 times larger than *HST*. The stars are 100 times fainter but we have 100 times the collecting area, so we get the same signal. The sky background is 100 times brighter (per unit area on the sky, due to the larger collecting area), but the area of each resolution element is 100 times smaller, so the sky signal within a resolution element stays the same. Thus, the SNR is the same for a given observing

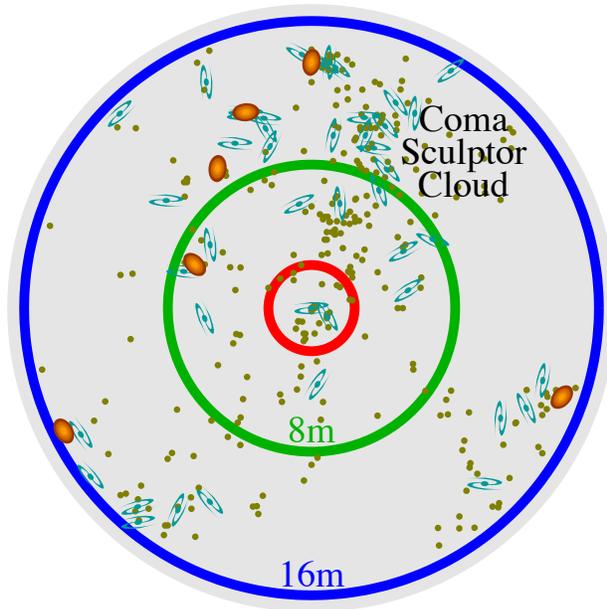

**Fig. 3 -** Giant spirals (blue symbols), giant el-
lipticals (red symbols), and dwarfs (brown
symbols) within 12 Mpc of the Milky Way
(center), deprojected to show actual distances.
Concentric circles indicate where space obser-
vatories can obtain SNR=5 photometry of a
solar analog star with 100 hours of observa-
tions split between two optical bands, thus ob-
taining the star formation history. The observa-
tories indicated are *HST* (red circle), *ATLAST*
8m (green circle), and *ATLAST* 16m (blue cir-
cle).

time. Surface brightness is conserved - a given patch of stars is in an area 100 times smaller, but
the stars are 100 times fainter and we are putting 100 times more resolution elements there. If
the larger telescope has the same field of view as the smaller telescope, the larger telescope will
have the advantage of sampling more physical real estate in the more distant galaxy. Further-
more, we are comparing here the performance of differently sized observatories at their distant
limits, but the larger telescope provides enormous advantages in nearby galaxies that are within
reach of both. The larger telescope can not only probe more crowded regions (instead of the faint
outskirts currently accessible to *HST*) but can also survey much more efficiently, because the ex-
posure time to obtain background-limited photometry of stars at a fixed distance is inversely
proportional to the fourth power of aperture size.

**Future Capabilities Needed**

With an observatory similar to *HST* having an aperture in the 8 to 16 meter range, we can finally
explore the full range of galaxy types and the variety of their structures, because the reach of
such a telescope extends well beyond our rural Local Group into the more cosmopolitan Coma
Sculptor Cloud (Fig. 3). The *Advanced Technology Large Aperture Space Telescope (ATLAST)*
would be such a telescope, and it is currently the subject of a NASA-funded study led by M.
Postman. An 8 meter aperture is needed to reach at least one giant elliptical, while a 16 meter
aperture is needed to reach a significant sample of both giant ellipticals and giant spirals (Fig. 4).

In the current era, there is a synergy between *HST* and large ground telescopes like Keck. *HST*
imaging can provide accurate photometry of faint dwarf stars at V~30 mag in Local Group gal-
axies, while Keck spectroscopy can provide radial velocities for bright giant stars at V~22 mag
in the same populations, providing important kinematical context (e.g., see Fig. 1). The James
Webb Space Telscope (*JWST* ) will extend our reach for this work to galaxies 50% more distant
than those available to *HST*. In the era of a 16m *ATLAST* and a 30m ground telescope (e.g.,

TMT), this synergy will move outward to much more distant galaxies, with *ATLAST* obtaining photometry of V~35 mag dwarf stars in the Coma Sculptor Cloud and TMT obtaining kinematics of bright giants in the same populations. These faint dwarf stars are effectively impossible with TMT, requiring Gigaseconds of integration even for an isolated star. A space platform is required for this type of work, because one needs stable high-precision photometry for thousands of stars in large crowded large fields with faint sky backgrounds, and this photometry must be accurate for stars spanning a large dynamic range (~$L_{Sun}$ to 10,000 times brighter).

## References

Bovill, M. S., & Ricotti, M. 2009, ApJ, 693, 1859

Brown, T.M., et al. 2003, ApJ, 592, L17

Brown, T.M., et al. 2006, ApJ, 652, 323

Brown, T.M., et al. 2008, ApJ, 658, L121

Brown, T.M., et al. 2012, ApJ, 753, L121

Bullock, J., & Johnston, K. 2005, ApJ, 635, 931

Cole, A.A., et al. 2007, ApJ, 659, L17

Dolphin, A.E. 2002, MNRAS, 332, 91

Durrell, P.R., et al. 1994, AJ, 108, 2114

Durrell, P.R., et al. 2001, AJ, 121, 2557

Fardal, M.A., et al. 2007, MNRAS, 380, 15

Font, A., et al., 2008, ApJ, 673, 215

Gallart, C., et al. 1999, AJ, 118, 2245

Gilbert, K.M., et al. 2007, ApJ, 668, 245

Guhathakurta, P., et al. 2005, astro-ph/0502366

Hammer, F., et al. 2007, ApJ, 662, 322

Harris, J., & Zaritsky, D. 2001, ApJS, 136, 25

Holtzman, J.A., et al. 1999, AJ, 118, 2262

Ibata, R.A., et al. 1994, Nature, 370, 194

Ibata, R.A., et al. 2001, Nature, 412, 49

Irwin, M.J., et al. 2005, ApJ, 628, L105

Kalirai, J.S., et al. 2006, ApJ, 648, 389

Majewski, S.R., et al. 2007, ApJ, 670, L9

Marinoni, C., et al. 1999, ApJ, 521, 50

Moore, B., et al. 1999, ApJ, 524, L19

Mould, J., & Kristian, J. 1986, ApJ, 305, 591

Sandage, A.R. 1953, AJ, 58, 61

Tosi, M., et al. 1991, AJ, 102, 951

Tumlinson, J. 2010, ApJ, 708, 1398

van den Bergh, S. 2000, PASP, 112, 529

Willman, B., et al. 2005, ApJ, 626, L85 2005

Zucker, D.B., et al. 2006, ApJ, 643, L103

Zucker, D.B., et al. 2007, ApJ, 659, L21

**Fig. 4 -** The cumulative number of galaxies where space telescopes can obtain the detailed star formation history, as a function of aperture diameter, assuming 100 hours of observations split between the V & I bands. The star formation history can be measured anywhere one can obtain SNR=5 photometry of solar analog stars. Representative observatories are indicated by grey lines. To measure the star formation history in a significant sample of giant elliptical galaxies, one needs at least a 16m space telescope.

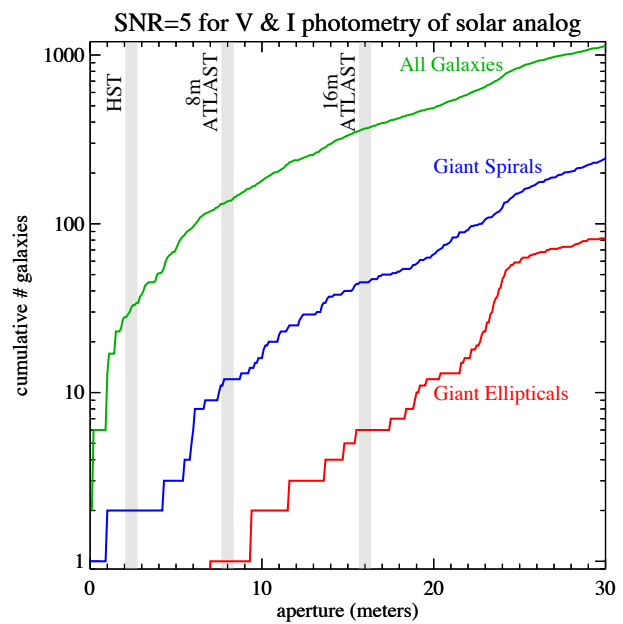